# The solubility of indium in liquid gallium supercooled to 12 K

Xiangyu Yin [a] and Gary S. Collins [b]

Department of Physics and Astronomy, Washington State University, Pullman, WA, USA

[a]desmond2046@gmail.com, [b]collins@wsu.edu

**Keywords:** solubility, metastable, supercooling, gallium, perturbed angular correlation, hyperfine interaction, nuclear quadrupole interaction, site preference.

**Abstract.** The method of perturbed angular correlation (PAC) was used to determine lattice locations of [111]In impurity probe atoms present in extreme dilution in the intermetallic compound $FeGa_3$. In slightly Ga-poor samples, probes were found to strongly prefer one of two inequivalent Ga-sites. In slightly Ga-rich samples at room temperature, 293 K, the PAC spectrum exhibited an unperturbed quadrupole interaction signal that is consistent with indium probes dissolved in small liquid pools of the excess Ga. A myriad of such pools are probably located along grain boundaries in the sample. Cooling from 293 K down to 12 K, the site fraction of indium in liquid decreased, being offset by the increase in a signal attributed to indium solutes in precipitates with other impurities at the sides of the Ga pools. However, these changes were completely reversible upon heating, and no crystallization of the liquid gallium pools was observed down to 12 K. This is attributed to the extraordinarily small volumes for the pools, which, while not measured directly, are orders of magnitude smaller than cubic microns. The measured temperature dependence of the site fraction of indium in the liquid was used to extend the metastable solubility curve for indium in liquid gallium down to a temperature of 150 K, much lower than the eutectic temperature of Ga-In at 288.5 K.

## Introduction

Liquid gallium normally solidifies at 30°C and 1 atm into α-gallium, which has a complex, orthorhombic crystal structure. However, like many other metals, gallium can be cooled and remain in a metastable liquid state below the solidification point [1] that is called supercooled. In general, heterogeneous nucleation of the solid phase by nucleating centers can be reduced by increasing the purity of the liquid and reducing the size of the liquid drops [1,2]. For small enough drop sizes, the probability of having an extrinsic nucleating center is zero, so that only homogeneous nucleation is possible. Typically, metals can be supercooled to 80% of their melting temperatures when volumes of liquid "pools" have been reduced to in the range $10^3 - 10^6$ cubic microns [1]. Supercooled liquid gallium has been reported to freeze at 265 K when in 16 nm pores [3], at ~215 K when in 8 nm pores [2] and has been reported to remain liquid down to 34 K in small droplets [4].

In this laboratory, there is an active research program studying diffusion in intermetallic compounds using a nuclear hyperfine interaction technique, perturbed angular correlation of gamma rays (PAC). A second area of research has been to determine and understand the systematics of site preferences taken up by dilute impurities in intermetallic compounds. These studies have been carried out through measurements of nuclear quadrupole interactions of [111]In/Cd impurity PAC probe atoms. Using PAC, one can measure electric field gradients at nuclei of the probe atoms. Lattice locations of the probes can be identified when the different sites occupied have different point symmetries (see, e.g., [5, 6]). Diffusional jumps of the probe atoms between sites having different symmetries leads to relaxation of the quadrupole interaction that can be fitted to obtain the jump frequency (see, e.g., [7,8]). It has been found that both jump frequencies and site preferences of dilute impurities are highly sensitive to precise compositions of compounds, even in "line compounds" for which widths of the phase fields are less than 1 at.% [9]. Consequently, our normal procedure for studying binary line compounds has been to make measurements on a pair of samples, with each prepared to have a slight excess of each element. Impurity probe atoms can switch sites due to even small changes in composition. For example, [111]In impurities in the cubic Laves phase $GdAl_2$ occupy Gd-sites at the Gd-poor boundary composition, but Al-sites at the Al-poor boundary [10]. A general rule of thumb is that an impurity tends to occupy sites of the element in which there is a deficiency because this reduces the overall defect count. However, one must allow for the possibility that the equilibrium location of an impurity may not be any of the crystallographic sites. Thus, for example, while [111]In probes in $Ni_2Al_3$ occupy one of two inequivalent Al-sites in Ni-rich samples, in Al-rich samples they are "ejected" to non-lattice sites such as grain boundaries that exhibit inhomogeneously broadened, quadrupole interactions [5]. Thus, one must consider noncrystallographic "lattice sinks" as possible preferred sites.

Recent studies were carried out to measure jump frequencies of [111]In probes in the tetragonal gallium-based intermetallic compound $FeGa_3$ [11]. For Ga-poor $FeGa_3$, the probes were found to occupy one of two inequivalent Ga-sites with well-defined quadrupole interactions. For Ga-rich $FeGa_3$ samples, the exhibited behavior was not unlike that observed for $Ni_2Al_3$. PAC measurements at 293K exhibited a large site fraction of an unperturbed signal typical of PAC probes dissolved in a liquid. In liquids, transient electric field gradients (EFG) caused by molecular collisions are motionally averaged to zero EFG. An unperturbed signal would arise instead if probe atoms are at cubic sites, but there are no such sites in $FeGa_3$ or other Ga-rich intermetallic phases. In any case, x-ray analysis showed only the $FeGa_3$ structure. Thus, it is concluded that, just as in Al-rich $Ni_2Al_3$, indium in Ga-rich $FeGa_3$ gets expelled from the crystal structure, here becoming dissolved in small liquid pools of excess gallium that are probably located at grain boundaries. The eutectic temperature of Ga-In is 288.5K, below which temperature all gallium and indium is solid in the equilibrium state. Thus, the equilibrium state of the sample at 293 K has a large volume fraction of slightly Ga-rich $FeGa_3$ that contains no probe atoms, a small amount of solid alpha-Ga due to crystallization of a few pools, and many small pools of Ga containing most of the In in solution.

The total volume fraction of all pools is very small. The [111]In activity is purchased in carrier-free form [12], and only an activity of 10 microcuries is introduced into a sample for the measurements, corresponding to $10^{10}$ atoms. The total volume of each sample was about 0.015 cm$^3$, or about $10^{21}$ atoms. The eutectic composition is of the order of 10 at.% In [13], so that the volume fraction of the liquid Ga-In pools is about $10^{-10}$, and the total volume is about 2 cubic microns. In order for essentially all [111]In probe atoms to be able to diffuse to the liquid pools, there must be very many pools. Thus, volumes of the individual pools are orders of magnitude smaller than a cubic micron.

Measurements between 293K and 12K showed reversible changes and no evidence of crystallization of the liquid pools. Over that temperature range, the site fraction of indium in the supercooled liquid decreased with temperature owing to a decreasing solubility of indium. The measurements of the site fraction in indium in the supercooled liquid are used to extend the solubility curve for indium to low temperature.

**Experiments**

Samples were made by arc-melting 99.99% pure Fe and 99.99999% pure Ga with [111]In under argon in an arc-furnace. Measurements were made using a standard four-counter PAC spectrometer [10] with samples in a temperature-controlled closed-cycle He refrigerator. Space precludes a detailed description of PAC spectroscopy; see [10, 14] for details beyond the level presented here. $FeGa_3$ is a line compound that has the tetragonal structure shown in Fig. 1 [15]. The PAC spectrum measured for a Ga-poor sample is shown in Fig. 2, exhibiting a unique quadrupole interaction signal that is attributed to probe atoms on Ga-sites at corners of the squares shown in Fig. 1 [11]. The lack of inhomogeneous broadening in Fig. 2 is typical of quadrupole interactions at probe atoms in highly-ordered defect-free compounds.

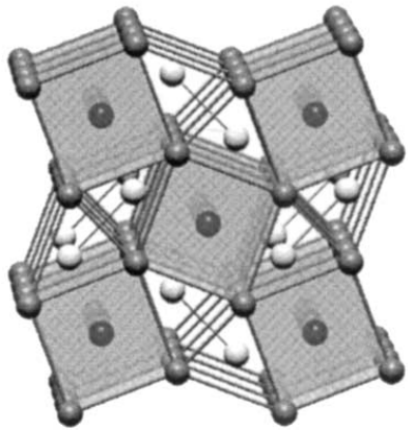 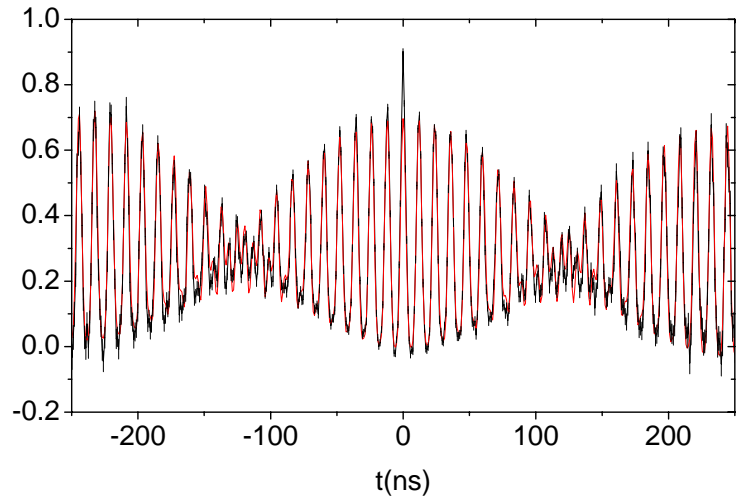

Fig. 1 Crystal structure of FeGa3 [15].    Fig. 2. PAC spectrum of Ga-poor FeGa3 at 293 K.

PAC spectra measured on Ga-rich samples are shown in Fig. 3. They do not exhibit any of the signal observed in Fig. 2 for probes on one of the two Ga-sites, although x-ray measurements confirmed that the samples had the FeGa$_3$ structure. Instead, site fractions of three other signals were observed over the range from 293K down to 12K: (1) an unperturbed signal due to indium dissolved in liquid gallium; (2) a broadly distributed quadrupole interaction that appears as a "glitch" near time zero, in the middle of the figure, and that is attributed to indium atoms precipitating on the sides of the liquid pools with other impurities; and (3) a small, temperature-independent site fraction of a static, high-frequency quadrupole interaction for isolated indium impurities in solid, alpha-gallium [16]. With decreasing temperature, the site fraction of the liquid, being offset by an increase in the site fraction of the glitch signal. Changes with temperature were completely reversible down to the lowest measurement temperature, 12K, with no evidence of crystallization of the liquid gallium into alpha-gallium crystals containing indium impurities, or of crystallization of indium into indium nanocrystals.

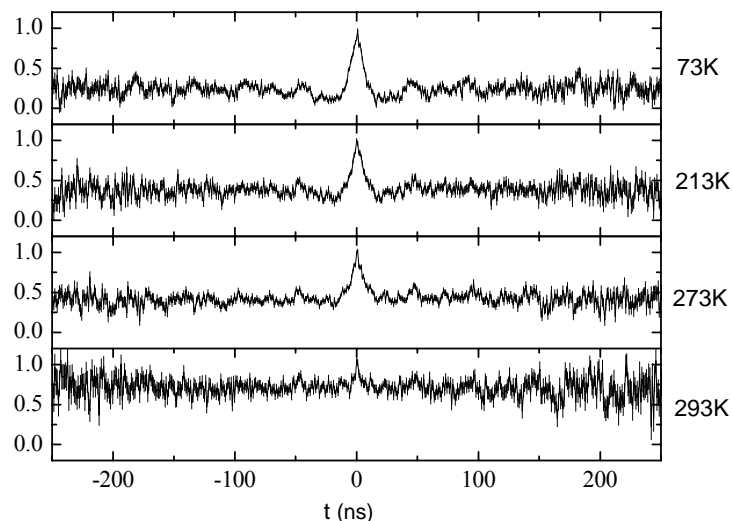

Fig. 3. PAC spectra of Ga-rich FeGa$_3$ measured at the indicated temperatures. With decreasing temperature, the unperturbed signal (vertical offset) for indium in liquid Ga is replaced by an inhomogeneously broadened signal, visible as the "glitch" near time zero. The small, high-frequency signal observed in all four spectra is due to indium probes incorporated in pools of gallium that have solidified into alpha-Ga.

Fig. 4 shows the temperature dependence of the site fraction of $^{111}$In probe atoms in the liquid phase. The decrease in the liquid site fraction is offset by an increase in an inhomogeneously broadened, static quadrupole interaction, which is attributed to precipitation of indium along walls of the pools with other impurities as the solubility limit is exceeded. In addition, there is a 10% temperature-independent site fraction for indium in alpha-Ga, which has a high frequency quadrupole interaction. (The alpha-Ga signal is attributed to solidification of a small fraction of the liquid pools, and is irreversible below the melting temperature of 303 K.) Data shown in Fig. 4 were obtained using two samples, with orders of measurements shown in Table 1. From the order of measurements, it can be seen that the measurements are reversible, showing that there was no crystallization of the liquid pools during cooling down to 12 K.

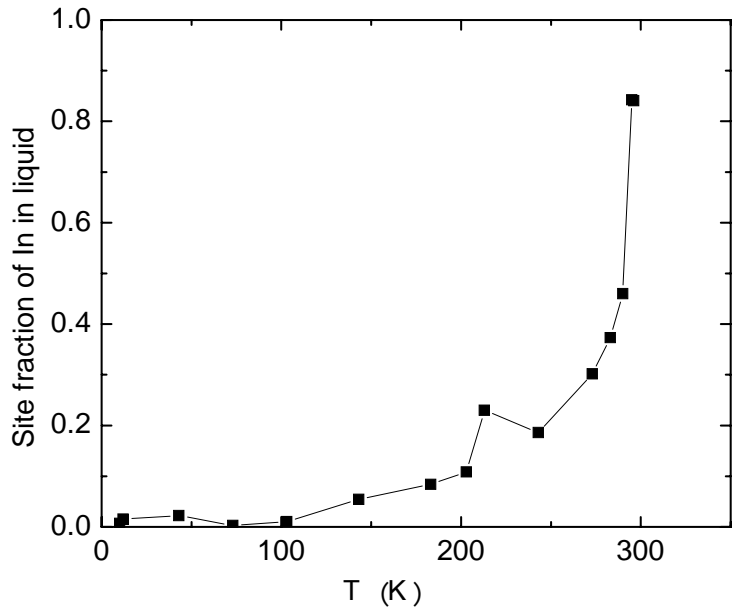

Fig. 4. Site fraction of In probe atoms in liquid Ga-In.

| Sample | Order of measurement | T(K) |
|---|---|---|
| 1 | 1 | 296 |
|   | 2 | 73 |
|   | 3 | 273 |
|   | 4 | 143 |
|   | 5 | 295 |
|   | 6 | 213 |
|   | 7 | 283 |
|   | 8 | 290 |
|   | 9 | 12 |
|   | 10 | 183 |
|   | 11 | 103 |
| 2 | 1 | 243 |
|   | 2 | 203 |
|   | 3 | 43 |
|   | 4 | 10 |

Table 1. Orders of measurement for data shown in Fig. 4.

### Solubility curve for indium in supercooled gallium liquid

The solubility curve for indium in liquid gallium can be determined from the data down to 12 K shown in Fig. 4, subject to normalizing the solubility at one temperature. This was done by normalizing the site fraction at the eutectic temperature of Ga-In alloys, 288.5K, to the eutectic composition of 14.2 at.% In [13]. Using this normalization, terminal solubilities of indium in gallium calculated at other temperatures are listed in Table 2. These data are plotted in Fig. 5 on top of a metastable phase diagram for Ga-In from ref. [13].

| T (K) | Terminal Solubility (at. %) |
|---|---|
| (288.5K) | (14.2) |
| 283 | 11.5 |
| 273 | 9.3 |
| 243 | 5.8 |
| 203 | 3.3 |
| 143 | 1.7 |

Table 2. Solubilities of indium in gallium liquid determined from data in Fig. 4.

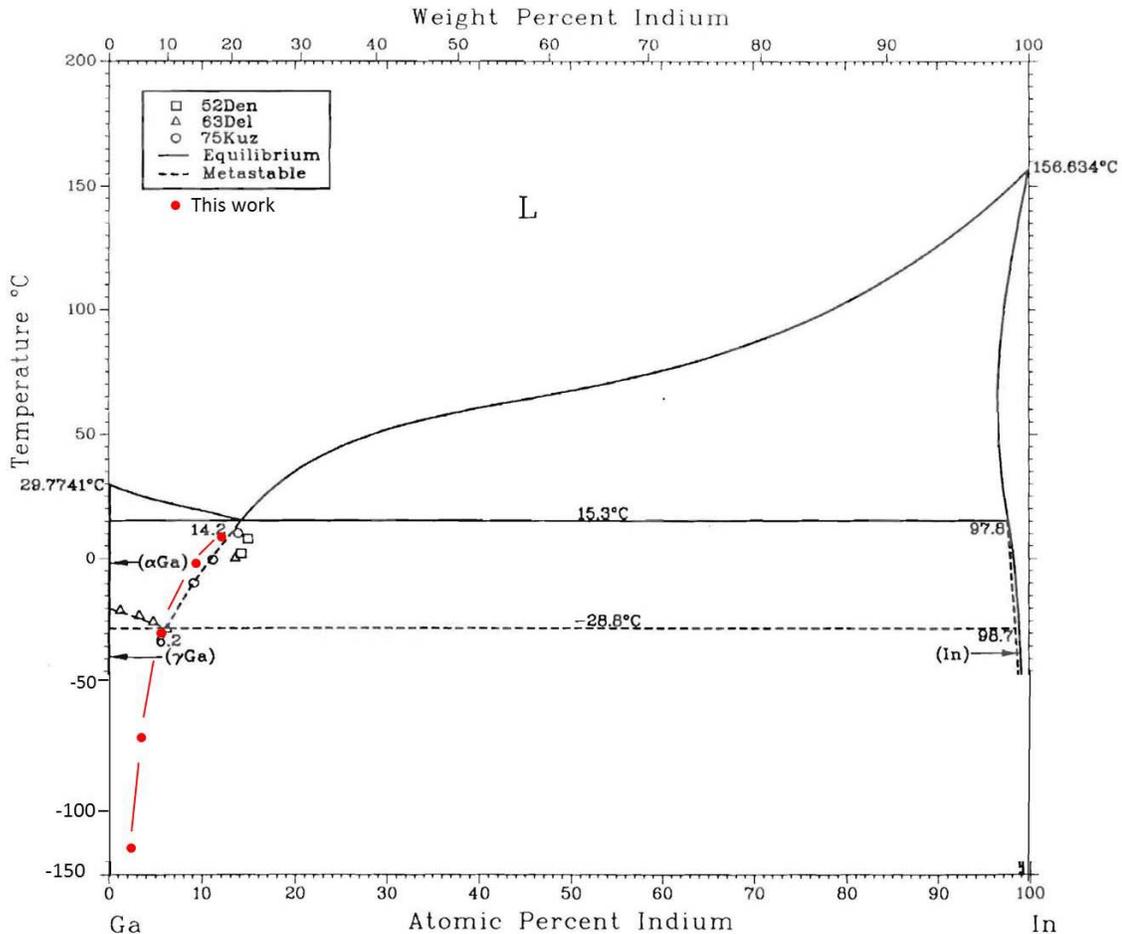

Fig. 5. Metastable binary phase diagram of Ga-In, reproduced from ref. [13], with addition of data points giving the terminal solubility of indium in supercooled gallium obtained in this work.

The phase diagram [13] in Fig. 5 shows equilibrium phase boundaries (solid lines without data points), three sets of data measured previously for metastable, supercooled liquid gallium (dashed lines and open symbols), and measurements from this work (solid circles with connecting lines. While the eutectic temperature for supercooled gallium shown in the diagram [13] was 244.4K, the eutectic temperature in the present measurements was below ~12 K and could not be measured.

**Discussion**

**Nonobservation of homogeneous nucleation**. The suppression of heterogeneous nucleation in the present experiments is attributed to very small sizes of the liquid pools, so that few pools held impurity crystallization nuclei [1]. However, there was also no evidence of homogeneous nucleation down to 12K. Strong reduction in collision rates in the liquid at low temperatures could impede homogeneous nucleation that occurs through random creation of crystallization nuclei.

**Site preferences of indium impurities**. $^{111}$In PAC probes strongly prefer not to be surrounded by Al-atoms or Ga-atoms, as they are at the Ni-site in $Ni_2Al_3$ [5], or at the Fe-site in $FeGa_3$, in the present work. Thus, in Ga-rich $FeGa_3$ samples there is strong probe segregation to gallium liquid. In additional measurements to be reported elsewhere [17], lattice locations of $^{111}$In probes were also studied in 8N pure gallium. No supercooling was observed in bulk gallium, and most $^{111}$In remained in liquid Ga-In above the eutectic temperature of 288.5K. Below the eutectic temperature, indium was observed to solidify and grow into tetragonal indium crystals, as expected, with a small fraction of $^{111}$In incorporated into solid alpha-Ga. The terminal solubility of indium in alpha-Ga was determined to be close to a mole fraction of $10^{-12}$, a remarkably low value [17].

**Application of this approach to other supercooled metals.** In principle, terminal solubilities of PAC probes could be measured in other supercooled metals. One needs a combination of a solid host element and binary compound in both of which the probe has exceptionally small solubilities, and the host element and compound must have no intermediate phases between their compositions..

**Summary and acknowledgment**

PAC experiments were carried out on extremely dilute concentrations of $^{111}$In probe impurities in Ga-rich samples of the intermetallic FeGa$_3$. At 293K, all probes were excluded from the compound and determined to be mostly in small pools of supercooled gallium metals that were probably located in grain boundaries of the compound. The site fraction of indium in the liquid decreased with decreasing temperature, due to a decreasing solubility. The temperature dependence of the liquid site fraction was found to be completely reversible after cooling to as low as 12K, demonstrating that the pools remained supercooled, without crystallization, down to that temperature. The measurements were used to extend the solubility curve for indium in liquid gallium down to ~150K.

This work was supported in part by the National Science Foundation under grant DMR 09-04096 (Metals Program).

**References**

[1] V.P. Skripov, *Metastable Liquids* (John Wiley, New York, 1973) pp. 136-142

[2] B. F. Borisov et al., J. Physics: Condens. Matter 11 (1999) pp. 10259–10268.

[3] E.V. Charnaya et al., Phys. Rev. B 72 (2005) p. 35406.

[4] A. Di Cicco et al., Europhys. Letters 27 (1994) pp. 407-412.

[5] Matthew O. Zacate and Gary S. Collins, Physical Review B70 (2004) 24202.

[6] John P. Bevington, Farida Selim and Gary S. Collins, Hyperfine Interact. 177 (2007) pp. 15-19.

[7] M.O. Zacate, A. Favrot and G.S. Collins, Phys. Rev. Lett. 92 (2004) 22590; Gary S. Collins et al., Hyperfine Interact. 159 (2005) p. 1; Gary S. Collins et al., Phys. Rev. Lett. 102 (2009) 155901.

[8] Egbert R. Nieuwenhuis, Matthew O. Zacate and Gary S. Collins, Defect and Diffusion Forum 264 (2007) 27-32.; Farida Selim, John P. Bevington and Gary S. Collins, Hyperfine Interact. 178 (2007) 87-90.

[9] Gary S. Collins, Journal of Materials Science 42 (2007) 1915-1919.

[10] Matthew O. Zacate and Gary S. Collins, Physical Review B69 (2004) 174202.

[11] Randal Newhouse, Prastuti Singh and Gary S. Collins (to be published).

[12] Perkin-Elmer, catalog number NEZ-304.

[13] C.E.T. White and H. Okamoto, eds. *Phase diagrams of indium alloys and their engineering applications* (ASM International, Materials Park, Ohio, 1992) p. 113.

[14] Günter Schatz and Alois Weidinger, *Nuclear Condensed Matter Physics* (Wiley, 1995).

[15] Ulrich Häussermann et al., J. Solid State Chemistry 165 (2002) pp. 94-99.

[16] M. Menningen, H. Haas and H. Rinneberg, Phys. Rev. B 35 (1987) 8378.

[17] Xiangyu Yin and Gary S. Collins (to be published).